\documentclass[aps,prd,twocolumn,superscriptaddress,nofootinbib,groupedaddress,10pt]{revtex4-2}
 
\pdfoutput=1

\usepackage{slashed}
\usepackage{color,colortbl,verbatim}
\usepackage{latexsym}
\usepackage{amssymb}
\usepackage{amsmath}
\usepackage{graphicx}
\usepackage[dvipsnames]{xcolor}
\usepackage{adjustbox}

\usepackage{array}
\usepackage{adjustbox}
\usepackage{booktabs}
\usepackage{multirow}
\usepackage{rotating}
\usepackage{soul}
\usepackage{mathtools}
\usepackage{calrsfs}
\usepackage{makecell}

\usepackage{hyperref}

\definecolor{LightCyan}{rgb}{0.88,1,1}

\newcommand{\ii}{\ensuremath{\mathrm{i}}}

\DeclareMathOperator*{\sumint}{
\mathchoice
 {\ooalign{$\displaystyle\sum$\cr\hidewidth$\displaystyle\int$\hidewidth\cr}}
 {\ooalign{\raisebox{.14\height}{\scalebox{.7}{$\textstyle\sum$}}\cr\hidewidth$\textstyle\int$\hidewidth\cr}}
 {\ooalign{\raisebox{.2\height}{\scalebox{.6}{$\scriptstyle\sum$}}\cr$\scriptstyle\int$\cr}}
 {\ooalign{\raisebox{.2\height}{\scalebox{.6}{$\scriptstyle\sum$}}\cr$\scriptstyle\int$\cr}}
}

\newcommand{\blue}[1]{\textcolor{blue}{#1}}

\begin{document}

\title{Hot news on the phase structure of the SMEFT}

\author{Mikael Chala}
\email{mikael.chala@ugr.es}
\author{Maria Cristina Fiore}
\email{mcristinafiore@ugr.es}
\author{Luis Gil}
\email{lgil@ugr.es}

\affiliation{Departamento de F\'isica Te\'orica y del Cosmos,
Universidad de Granada, E--18071 Granada, Spain}

\begin{abstract} 
We perform dimensional reduction of the dimension-six SMEFT to order $g^4$ in coupling constants $g$. This analysis includes one-loop contributions to kinetic terms and quartic couplings; as well as two-loop contributions, where operators such as four-fermion interactions first appear, to squared mass terms. Using lattice data, we provide evidence that, in contrast with previous statements in the literature, the SMEFT may undergo a first-order phase transition even for null $|\phi|^6$ at zero temperature ---where $\phi$ is the Higgs doublet--- opening the door to entirely unexplored directions in model building. In such case, however, our analysis implies that the phase transition is not sufficiently strong, narrowing down the new physics that allows for electroweak baryogenesis.
\end{abstract}

\maketitle

\section{Introduction}

The modern approach to the study of high-temperature equilibrium phenomena within quantum field theory is based on the formalism of \textit{dimensional reduction} (DR) \cite{Ginsparg:1980ef,Appelquist:1981vg}. This method relies on the construction of a static, 3-dimensional (3D) effective-field theory (EFT), %(3D EFT), 
exploiting the hierarchy of scales between the temperature, $T$, and the masses, $m$, of light fields. In practice, all heavy Matsubara modes \cite{Matsubara:1955ws}, with thermal masses of order $\pi T$, are integrated out in favor of a theory with light bosonic zero modes whose interactions encode all temperature dependence. Perturbative calculations in the 3D EFT framework can bypass numerous difficulties encountered within finite-temperature calculations in four dimensions (4D)~\cite{Lofgren:2023sep}. Moreover, close to a critical point, where non-Abelian gauge theories become non-perturbative \cite{Linde:1980ts}, the 3D dynamics can be simulated on the lattice~\cite{Farakos:1994xh,Kajantie:1995kf,Laine:1995np,Laine:1997dy,Gurtler:1997hr,Rummukainen:1998as,Laine:1998jb,Moore:2000jw,Arnold:2001ir,Sun:2002cc,DOnofrio:2015gop,Gould:2022ran,Annala:2025aci}.

For the Standard Model (SM), this 3D EFT is fully known to order $g^4$~\cite{Kajantie:1995dw}, where $g \sim Y \sim \sqrt{\lambda}$ stand for the $SU(2)$, Yukawa and quartic couplings. (The full one-loop $\mathcal{O}(g^6)$ corrections have been computed in Ref.~\cite{Chala:2025aiz}; see also Ref.~\cite{Moore:1995jv}.) The phase structure of this theory was successfully understood on the lattice, with the remarkable finding that no first-order phase transition (FOPT) occurs within the electroweak sector of the SM \cite{Kajantie:1996mn, Gurtler:1997hr, Csikor:1998eu}. This result, together with the future prospect of observing its imprint on a gravitational wave background \cite{Harry:2006fi,Kawamura:2006up,Ruan:2018tsw,LIGOScientific:2014pky,Caprini:2019egz}, has motivated an extensive exploration of physics beyond the SM that can allow for a FOPT \cite{Brauner:2016fla,
Andersen:2017ika,Niemi:2018asa,Gorda:2018hvi,Kainulainen:2019kyp,Croon:2020cgk,Gould:2019qek,Niemi:2020hto,Gould:2021ccf,Gould:2021dzl,Schicho:2021gca,Niemi:2021qvp,Camargo-Molina:2021zgz,Niemi:2022bjg,Ekstedt:2022ceo,Gould:2022ran,Ekstedt:2022zro,Biondini:2022ggt,Schicho:2022wty,Lofgren:2021ogg,Gould:2023jbz,Kierkla:2023von,Aarts:2023vsf,Niemi:2024axp,Chala:2024xll,Qin:2024idc,Gould:2024jjt,Chakrabortty:2024wto,Niemi:2024vzw,Kierkla:2025qyz,Bhatnagar:2025jhh,LopezMiras:2025zak,Bernardo:2025vkz,Chala:2025aiz,Zhu:2025pht,Chala:2025oul,Li:2025kyo,Annala:2025aci,Bhatnagar:2025jhh,Navarrete:2025yxy}.

As an alternative to a case-by-case analysis of viable SM extensions, some studies have taken a model-independent approach and explored the electroweak PT (EWPT) within the SMEFT~\cite{Camargo-Molina:2021zgz, Camargo-Molina:2024sde, Chala:2025aiz},
which extends the SM with all gauge-invariant higher-dimensional operators suppressed by powers of an energy cutoff $\Lambda$
~\cite{Buchmuller:1985jz,Grzadkowski:2010es}.
This framework enables a systematic investigation of the conditions under which a FOPT can emerge, and their connection with experimental input from colliders~\cite{Huang:2015izx,Huang:2017jws,Brivio:2017vri,Ramsey-Musolf:2019lsf,Isidori:2023pyp,Aebischer:2025qhh}.

However, and despite the growing interest on the matter, little is yet known about the thermodynamics of the SMEFT~\cite{Chala:2025aiz}. In this article, we make a substantial leap forward in two fronts. First, we consider the full dimension-six SMEFT, including even operators that appear only at loop level in weakly-coupled ultraviolet (UV) completions of the SMEFT. And second, we compute the two-loop matching corrections of 3D mass terms. Altogether, we achieve for the first time full $\mathcal{O}(g^4)$ accuracy in the description of the SMEFT at high temperature~\footnote{For SMEFT interactions, we assume that the power counting is $c_{X^3} \sim g$, $c_{X^2 \phi^4} \sim c_{\phi^4 D^2}\sim c_{\psi^2 X \phi} \sim c_{\psi^2 \phi^2} \sim c_{\psi^4} \sim g^2$, $c_{\psi^2\phi^3}\sim g^3$ and $c_{\phi^6}\sim g^4$; where we indicated the number of gauge field strength tensors ($X$), scalars ($\phi$), fermions ($\psi$) and derivatives ($D$).}.

On the basis of this result and of up-to-date lattice data~\cite{Gould:2022ran}, we explore the possibility that the SMEFT undergoes a FOPT in regions of the parameter space previously overlooked in the literature. This sharply refines our understanding of the electroweak sector's thermal history.

\section{Calculation}
We use the following conventions for the SMEFT Lagrangian:
\begin{align}
 \mathcal{L}_\text{SMEFT} &= \mathcal{L}_\text{SM} + \frac{1}{\Lambda^2}\sum_i c_i\mathcal{O}_i + \mathcal{O}\left(\frac{1}{\Lambda^4}\right)\,,
\end{align}
where $c_i$ are Wilson coefficients (WC) and the sum runs over the non-baryon-non-lepton-violating dimension-six operators of the SMEFT, that we reproduce for completeness in Appendix \ref{app:operators} following the conventions of Ref.~\cite{Grzadkowski:2010es}.

\label{sec:calculations}

On the other hand, the SM Lagrangian reads:
\begin{align}
 \mathcal{L}_\text{SM} &= -\frac{1}{4} G_{\mu\nu}^A G^{A\mu\nu} - \frac{1}{4}W^I_{\mu\nu}W^{I\,\mu\nu}-\frac{1}{4}B_{\mu\nu}B^{\mu\nu} \\
 &+\overline{q}\ii\slashed{D}q+\overline{l}\ii\slashed{D} l+\overline{u}\ii\slashed{D}u+\overline{d}\ii\slashed{D}d+\overline{e}\ii\slashed{D}e\nonumber\\
 &+ (D_\mu\phi)^\dagger (D^\mu\phi) +\mu^2|\phi|^2 -\lambda|\phi|^4 \nonumber\\
 &- (\overline{q} \tilde{\phi} Y_u u + \overline{q}\phi Y_d d+\overline{l}\phi Y_e e + \text{h.c.})\,,\nonumber
\end{align}
where $u,d,e$ and $q,l$ represent the right-handed quarks and leptons and their left-handed counterparts, respectively. Likewise, $B$, $W$ and $G$ stand for the gauge bosons, with $g'$, $g$, $g_S$ being the corresponding gauge couplings. Finally, $\phi$ is the Higgs doublet, and $\tilde{\phi} = \ii\sigma_2\phi^*$, and we use the minus sign convention for the covariant derivative.

Since we are studying the EWPT in the SMEFT, we shall assume that the temperature scale is much larger than all relevant mass scales, but lower than the new physics scale, i.e. $\Lambda \gg T \gg m$. In this situation, the SMEFT expansion remains valid while allowing for the standard \cite{Gould:2023ovu} hierarchy of thermal scales:
\begin{equation}
    \underbrace{\pi T \vphantom{\left( \frac{g}{4 \pi} \right)} }_{\rm hard} \gg \underbrace{\left( \frac{g}{4 \pi} \right) \pi T}_{\rm soft} \gg \underbrace{\left( \frac{g}{4 \pi} \right)^{3/2} \pi T}_{\rm supersoft} \gg \underbrace{\left( \frac{g}{4 \pi} \right)^{2} \pi T}_{\rm ultrasoft}\,,
\end{equation}
written in terms of a weak coupling $g \ll 4 \pi$. 

\subsection{From hard to soft scale}

The first scale corresponds to the Matsubara modes of the SM fields, whose masses are of order $\pi T$. The second is associated to the screening of soft modes due to the hard-scale corrections, which induces a thermal mass for bosonic zero Matsubara modes. DR is the construction of the soft-scale 3D theory for zero Matsubara modes, and it amounts to integrating out all heavy non-zero modes. For the SM ---in the presence of SMEFT interactions up to order $g^4$--- this 3D EFT, in Euclidean space, reads as follows~\cite{Kajantie:1995dw}:
\begin{equation}\label{eq:3dlagrangian}
 \mathcal{L}_{\text{3}} = \mathcal{L}_{\text{3}}^{(m)} + \mathcal{L}_{\text{3}}^{(k)} + \mathcal{L}_{\text{3}}^{(\lambda)} + \mathcal{L}_{\text{3}}^{(\text{dim-6})}
\end{equation}
with
\begin{align}
&\mathcal{L}_{\text{3}}^{(m)} = m_\phi^2|\phi|^2 + \frac{1}{2}m_{B_0}^2 B_0^2 + \frac{1}{2}m_{W_0}^2 W_0^I W_0^I\,,
\end{align}
\begin{align}
 \mathcal{L}_{\text{3}}^{(k)} &= k_\phi(D_r\phi)^\dagger (D^r \phi)\nonumber\\
 &+ \frac{k_{B_0}}{2} (D_r B_0) (D_{r} B_0) + \frac{k_{W_0}}{2} (D_r W_{0}^{I})(D_{r} W_0^I)\nonumber\\
 &+\frac{k_B}{4} B_{rs} B_{rs} + \frac{k_W}{4}W_{rs}^I W_{rs}^I\,,
\end{align}
and
\begin{align}
 \mathcal{L}_{\text{3}}^{(\lambda)} &= \lambda_{\phi^4}|\phi|^4 + \lambda_{B_0^4} B_0^4 + \lambda_{W_0^4} (W_0^IW_{0I})^2\nonumber\\
 &+\lambda_{\phi^2 B_0^2} |\phi|^2 B_0^2 + \lambda_{\phi^2 W_0^2} |\phi|^2 W_0^I W_{0I} \nonumber\\
 &+ \lambda_{B_0^2 W_0^2}B_0^2 W_0^I W_0^I + \lambda_{\phi^2 B_0 W_0} B_0 \phi^\dagger \sigma^I\phi W_0^I\,.
\end{align}
We denote by $g_T$ and $g'_T$ the 3D gauge couplings of $SU(2)$ and $U(1)$, respectively. We do not include gluons in the 3D EFT, as they do not couple directly to the Higgs, and their effect on the EWPT is thus expected to be subleading. 

The last piece, $\mathcal{L}_{\text{3}}^{(\text{dim-6})}$, includes all dimension-six operators~\footnote{Note that we count energy dimensions in 4D units always.} involving the Higgs and the gauge bosons in the 3D theory. According to our power counting, at order $g^4$ these arise only at tree level from their 4D counterparts.
For brevity, we refrain from writing them here, as they will not be relevant for our results, but they can be found in Ref.~\cite{Chala:2025aiz}.

The parameters of the 3D EFT are computed upon requiring that Green's functions (i.e. off-shell correlators) in the 3D EFT reproduce the hard-region regime given by $Q^2 \sim (\pi T)^2 \gg P^2,\,\mu^2$ of those obtained in the (compactified) 4D theory \cite{Braaten:1995cm}. Here, $Q$ and $P=(0,\mathbf{p})$ represent loop and static external momenta, respectively.

To this aim, we use dedicated routines based on \texttt{FeynArts}~\cite{Hahn:2000kx} and \texttt{FeynCalc}~\cite{Shtabovenko:2016sxi}. We work in dimensional regularization with spacetime dimension $d=3-2\epsilon$ in the $\overline{\mathrm{MS}}$ scheme, and use arbitrary $R_\xi$ gauge. We only need the %following
one-loop
sum-integrals
%
%\begin{align}
 $I_{\alpha\beta\delta}^b \equiv \sumint_{Q} \frac{q_0^{2\beta} |\mathbf{q}|^{2\delta}}{Q^{2\alpha}}$
 and
 $I_{\alpha\beta\delta}^f \equiv \sumint_{\lbrace Q\rbrace} \frac{q_0^{2\beta} |\mathbf{q}|^{2\delta}}{Q^{2\alpha}} = (2^{2\alpha-2 \beta-2\delta-d}-1) I_{\alpha\beta\delta}^b\,,$
 %
%\end{align}
%
for $\beta \in \frac{1}{2} \mathbb{Z}$ and $\delta \in \mathbb{Z}$, with
\begin{align}\label{eq:sumintegrals}
 I_{\alpha\beta\delta}^b &= \frac{(e^{\gamma_E}\bar{\mu}^2)^\epsilon}{8\pi^2} \times \frac{1 + (-1)^{2\beta}}{2} \times (2\pi T)^{1+d-2(\alpha-\beta-\delta)} \nonumber\\ 
 &\times \frac{\Gamma(\alpha-\frac{d}{2}-\delta) \Gamma(\frac{d}{2}+\delta) \zeta (2(\alpha-\beta-\delta)-d)}{\Gamma(\frac{1}{2})\Gamma(\alpha)\Gamma(\frac{d}{2})} \,,
\end{align}
where $\overline{\mu}$ is the renormalization scale.

All our results turn out to be $\xi$-independent and finite. The first observation is the result of a trivial fact: all 3D parameters are physical at order $g^4$. It is only at higher-order that they receive corrections from 3D higher-dimensional operators which must be taken into account to achieve gauge independence in the physical basis~\cite{Chala:2025aiz,Bernardo:2025vkz}.

The second statement can be understood as follows in the massless Higgs limit ($\mu^2 = 0$). UV divergences are absorbed by the running of 4D renormalizable parameters, which are not renormalized by SMEFT operators~\cite{Jenkins:2013zja,Jenkins:2013wua,Alonso:2013hga}. IR divergences follow from the UV ones of the 3D EFT, but since 3D masses run only at two loops, SMEFT effects enter only at $\mathcal{O}(g^6)$~\footnote{Schematically, $\dot{m}_{\phi}^2 \sim \lambda_{\phi^4}^2$, but $\lambda_{\phi^4}\sim \lambda T + (c_\phi + \lambda c_{\phi\Box}+\cdots) T^2\,,$ where the ellipses stand for other SMEFT contributions. 
%Therefore, the running of $m_\phi^2$ scales with powers like $\lambda c_{\phi}$ or $\lambda^2 c_{\phi\Box}$, all of which are of order $g^6$.
}.

\subsubsection{One-loop results}\label{sec:one-loop-results}
At one loop, the non-vanishing SMEFT contributions to the 3D EFT 
%read
are those in Ref.~\cite{Chala:2025aiz} supplemented with the following highlighted terms, ensuing from operators that arise at loop level in UV completions of the SMEFT:
\begin{align}
    %
    %&\Delta k_\phi = \frac{1}{12}(c_{\phi D} - 2 c_{\phi\Box})T^2\,, \label{eq:matching_first}\\
    %
    &\Delta k_{B_0} = - \frac{2}{3} \blue{c_{\phi B}} T^2 \,, \label{eq:matching_first}\\
    &\Delta k_{W_0} = -\frac{2}{3} (\blue{c_{\phi W}}+3 g \blue{c_{3W}})T^2\,, \\
    &\Delta k_B = -\frac{2}{3} \blue{c_{\phi B}} T^2\,, \\
    &\Delta k_W = -\frac{2}{3} (\blue{c_{\phi W}}+3 g \blue{c_{3W}})T^2 \,, \\
    &\Delta m_\phi^2 = \frac{1}{12} \mu^2 \left( c_{\phi D} - 2 c_{\phi \Box}\right) T^2\,, \\
    &\Delta \lambda_{\phi^4} = \left[ -c_\phi + \frac{1}{4} \left( g'^2 \blue{c_{\phi B}} + g' g \blue{c_{\phi W B}} + 3 g^2 \blue{c_{\phi W}}\right) \right. \nonumber\\
    & \hphantom{\Delta \lambda_{\phi^4}} + \lambda c_{\phi \Box} + \frac{1}{48} \left( 3  (g'^2 + g^2) - 16 \lambda \right) c_{\phi D} \nonumber\\
    & \hphantom{\Delta \lambda_{\phi^4}} \left. - \frac{1}{12} \left( c_{e \phi} Y_e^* + 3 \left(c_{d \phi} Y_d^* + c_{u \phi} Y_u^*\right) + \rm{h. c.} \right) \right] T^3 \,. 
    \label{eq:matching_last}
\end{align}
Other couplings remain as in Ref. \cite{Chala:2025aiz}; also $m_\phi^2$ and $\lambda_{\phi^4}$, which we however show here for their relevance to the subsequent discussions.

\subsubsection{Two-loop results}\label{sec:two-loop-results}
We denote by $Q$ and $R$ the loop momenta.
Two-loop sum-integrals in our computation factorize into one-loop masters. Indeed, they arise from
$\frac{1}{\Lambda^2}\int \frac{d^4Qd^4R}{Q^{2\alpha}R^{2\beta}(Q-R)^{2\gamma}}$,
and dimensional analysis of squared mass terms requires scaling $\sim T^4/\Lambda^2$, hence $\alpha+\beta+\gamma=2$. This cannot be satisfied if all exponents are positive, precisely the case where factorization is nontrivial
~\footnote{If one of the exponents is negative, then one can expand the numerator to transform the sum-integral into a sum of factorized ones. For example: 
$$\sumint \frac{(Q-R)^2}{Q^4 R^4} = \frac{1}{2}\sumint \frac{Q^2+R^2-2Q\cdot R}{Q^4 R^4} = \frac{1}{2 Q^2 R^4}+\frac{1}{2 Q^4 R^2}\,;$$
where we have removed the term odd in $Q$ and $R$. Had $Q$ (or $R$) had the negative exponent, one could reach the same result upon changing variables $Q\to Q+R$. It is all this reasoning what we follow in this section.
}.

We propose the following steps to factorize the two-loop sum-integrals:
\begin{itemize}
 \item[(i)] In terms with $Q+R$, we use the identity $Q\cdot R = \frac{1}{2}\left[(Q+R)^2-Q^2-R^2\right]$. 
 \item[(ii)] In terms with both $Q+R$ and $Q$ ($R$) in the denominator, we perform the change of variables $Q+R\to R (Q)$. We note that $Q+R$ is bosonic if both $Q$ and $R$ have the same statistics, but it is fermionic otherwise.
 \item[(iii)] At this stage, no term involves $Q+R$ anymore. Then we expand products of loop variables using $Q\cdot R = q_0 r_0 + \bf{q}\cdot\bf{r}$.
 \item[(iv)] We set external Lorentz indices $(\mu,\nu)$ to $(0,0)$ or to $(i,j)$, depending on whether we match operators involving temporal or spatial components. We remove odd powers of $q_0,r_0,\bf{q}$ and $\bf{r}$.
 \item[(v)] We use tensor reduction, $q_i r_j \to \frac{1}{d} \boldsymbol{q} \cdot \boldsymbol{r} \delta_{ij}$, where $q_i$ ($r_j$) denotes the components of spatial loop momentum $\boldsymbol{q}$ ($\boldsymbol{r}$).
\end{itemize}
Finally, we evaluate all one-loop integrals using the formula in Eq.~\eqref{eq:sumintegrals}.

As an illustrative example, we explicitly compute the contribution of $c_{\phi e}$ to $m_\phi^2$ in the limit of vanishing $g', g$ and $\lambda$ and with only one family of fermions. To this aim, we consider the Green's function $\mathcal{G}_{\phi\phi}$ at zero momentum. We obtain:
\begin{align}
 \mathcal{G}_{\phi\phi} &\sim 4 \sumint_{\lbrace QR\rbrace} \left[\frac{1}{Q^2 (Q+R)^2} + \frac{Q\cdot R}{Q^2 R^2 (Q+R)^2}\right] \nonumber\\
 &\underset{(i)}{=} 2 \sumint_{\lbrace QR\rbrace} \left[ \frac{1}{Q^2 (Q+R)^2} + \frac{1}{Q^2 R^2}-\frac{1}{R^2 (Q+R)^2}\right] \nonumber\\
 &\underset{(ii)}{=} 2 \biggl[ \sumint_{\lbrace Q\rbrace R} \frac{1}{Q^2 R^2}  + \sumint_{\lbrace QR\rbrace}\frac{1}{Q^2 R^2} - \sumint_{Q\lbrace R\rbrace}\frac{1}{R^2 Q^2} \biggr]\,;
\end{align}
where $\sim$ indicates that we are ignoring the factor $c_{\phi e} |Y_e|^2$.
Substituting the sum-integrals and taking into account that, in the 3D EFT, $\mathcal{G}_{\phi\phi} = m_{\phi}^2$, we therefore have:
\begin{align}
m_{\phi}^2 & = -2 c_{\phi e} |Y_e|^2 \left[ I_{100}^b I_{100}^f + (I_{100}^f)^2 - I_{100}^f I_{100}^b\right]\nonumber\\
 &= -\frac{1}{288} c_{\phi e} |Y_e|^2 T^4 \,.
\end{align}

Now, including all dimension-six SMEFT operators, we obtain:
\begin{align}
\Delta m_{\phi}^2 &= \biggl[-\frac{1}{4} c_\phi + \frac{47}{3} g_S^2 c_{\phi G} + \frac{1}{576} |Y_u|^2 \bigl( 30 c_{\phi \Box} - 15 c_{\phi D} \nonumber \\ 
&+ 6 c_{\phi u} - 6 c_{\phi q}^{(1)} + 18 c_{\phi q}^{(3)} + 24 c_{q u}^{(1)} + 32 c_{q u}^{(8)} \bigr) \nonumber\\
&+ \frac{3}{64} \bigl( 16 g_S c_{u G} Y_u^* - 3 c_{u \phi} Y_u^* + \mathrm{h.c.} \bigr)\biggr] T^4 \,, \label{eq:mass_2loops}
\end{align}
where we have only written the limit in which $\mu^2,\lambda, g, g', Y_d, Y_e \to 0$ for the sake of brevity ($m^2_{B_0}$ and $m^2_{W_0}$ get no correction in this limit). We provide the full result, including both one- and two-loop matching corrections, in the ancillary file \texttt{matching.txt}.

From dimensional analysis, it can be also checked that one- and two-loop matching corrections to $m_\phi^2$ and $\lambda_{\phi^4}$ by dimension-8 SMEFT terms vanish.

\subsection{From soft to supersoft scale}
Temporal gauge bosons in the 3D EFT live at the soft scale, as they acquire positive squared masses proportional to $g^2 T^2$, which are parametrically larger than the thermal Higgs mass near the PT. Thus, temporal gauge bosons can be integrated out to build a supersoft-scale EFT where only the Higgs zero-mode and the spatial gauge bosons are present.

In good approximation, given that $g' \ll g$, the contribution of the $U(1)$ gauge bosons to the Higgs effective potential is subleading compared to that of the $SU(2)$ ones. This motivates the construction of a theory with $SU(2)$ + Higgs at the supersoft (SS) scale:
\begin{align}
 \mathcal{L}_{3, \text{SS}} &= \frac{1}{4}W_{rs}^I W^{I}_{rs} + (D_r\phi)^\dagger (D_r \phi) \nonumber \\
&+ m_3^2|\phi|^2 + \lambda_{3} |\phi|^4 + \mathcal{L}_{3, \text{SS}}^{(\text{dim-6})}\,,
\label{eq:SU2Higgs}
\end{align}
where we now dub $g_3$ the $SU(2)$ gauge coupling, and we again reuse the same names for the fields. $\mathcal{L}_{3, \text{SS}}^{(\text{dim-6})}$ encodes all dimension-six operators involving the $W$ and $\phi$ fields in $d=3$. These can also be read from Ref.~\cite{Chala:2025aiz}.

The matching onto this theory was computed to order $g^4$ in Ref.~\cite{Kajantie:1995dw}, in the absence of higher-dimensional operators, starting from the SM in the high-temperature limit. In this work, we shall use these results together with the contribution from SMEFT operators to the matching of the renormalizable parameters in the 3D EFT --Eqs.~\eqref{eq:matching_first} to \eqref{eq:matching_last}, and \eqref{eq:mass_2loops}--, and we will neglect the contribution from 3D higher-dimensional operators where justified.

\section{Phase structure}
\label{sec:phasestructure}
The phase structure of the theory described by the Lagrangian in Eq.~\eqref{eq:SU2Higgs} (in the absence of higher-dimensional operators) was studied on the lattice in terms of the dimensionless ratios $x\equiv \lambda_3/g_3^2$ and $y\equiv m_3^2/g_3^4$. As first determined in Ref.~\cite{Kajantie:1996mn}, the phase diagram presents a critical point located at $(x,y) = (x_*, y_*)$, with \cite{Gurtler:1997hr, Laine:1998jb}
\begin{equation}
    x_* = 0.0983(15)\,, \qquad y_* = -0.0175(13)\,,
\end{equation}
for renormalization scale $\overline{\mu}_3 = g_3^2$ in the 3D EFT, which we also use.

We can approximate the evolution of the SM and the SMEFT across this phase diagram using  $\mathcal{O}(g^2)$ matching in the limit of vanishing $g'$, together with generic deformations due to SMEFT interactions; see Appendix \ref{app:illustration}. A rough estimation indicates that new physics mainly need to introduce a moderately negative correction to $x$ (through a negative modification of $\lambda_3$ with respect to the SM value) in order to achieve a FOPT. On the basis of Eqs.~\eqref{eq:matching_first} to \eqref{eq:matching_last} and for concreteness, we focus on four particular operators in the SMEFT that can induce such shifts in $\lambda_3$: $\mathcal{O}_{t \phi}$, $\mathcal{O}_{\phi G}$, $\mathcal{O}_\phi$ and $\mathcal{O}_{\phi \Box}$.

The first is somewhat constrained by Higgs data~\cite{Ellis:2020unq,Ethier:2021bye,deBlas:2025xhe}, though its bounds depend significantly on the second, $\mathcal{O}_{\phi G}$. Using \texttt{SMEFiT}~\cite{Giani:2023gfq}, we effectively find~\footnote{We marginalize over all the WCs that do \textit{not} enter into Eqs.~\eqref{eq:matching_first} to \eqref{eq:matching_last} in the limit of $g',g,\lambda,Y_e,Y_d\to 0$. (Powers of these SM parameters are so small that, together with the loop suppression, not even large values of the marginalized WCs affect significantly the matching conditions). All others are set to zero.} $-6 \lesssim c_{t\phi}/\text{TeV}^2 \lesssim -2$;
see Appendix \ref{app:constraints}. We note that $\mathcal{O}_{t \phi}$ does not only affect $\lambda_3$ through its contribution to the matching, but also through a modification of the tree-level, zero temperature, relation between $Y_t$ and the physical parameters. On the other hand, $\mathcal{O}_{\phi G}$ itself does not modify the PT, both because it only enters $m_3^2$ (at two-loop level), and because its value is tightly constrained.

The bounds on $\mathcal{O}_{t \phi}$ are such that it cannot allow for a FOPT alone, but an additional contribution to $\lambda_3$ from different relatively unconstrained operators, such as $\mathcal{O}_\phi$ or $\mathcal{O}_{\phi \Box}$, is needed. The first of the two is not experimentally constrained, so we will take a conservative range of $[-1,1]$ TeV$^{-2}$. On the other hand, in the plane of $c_{t\phi}$ and $c_{\phi\Box}$ upon marginalization, the latter is bounded as $-2 \lesssim c_{\phi\Box}/\text{TeV}^2 \lesssim 2$; see Appendix \ref{app:constraints}.
These operators further modify the tree-level relation between $\mu^2$ and $\lambda$ and the physical parameters $v \sim 246$ GeV and $m_H\sim 125$ GeV; see e.g. Ref.~\cite{Postma:2020toi}. For instance, the 4D Higgs quartic coupling is modified as
\begin{equation}
    \lambda = \frac{m_H^2}{2 v^2} + \frac{3}{2} c_\phi v^2 - 4 c_{\phi \Box} \frac{m_H^2}{2}\,.\label{eq:lambda}
\end{equation}
Therefore, on top of the thermal contribution to $\lambda_3$ in Eq. \eqref{eq:matching_last}, the tree-level piece $\lambda_3\bigr|_\mathrm{tree} = \lambda T$ gets further corrected by dimension-six operators. We find these modifications to be often numerically as (and sometimes more) relevant as those from matching equations.
 
\begin{figure}[t]
    \centering
    \includegraphics[width=\columnwidth]{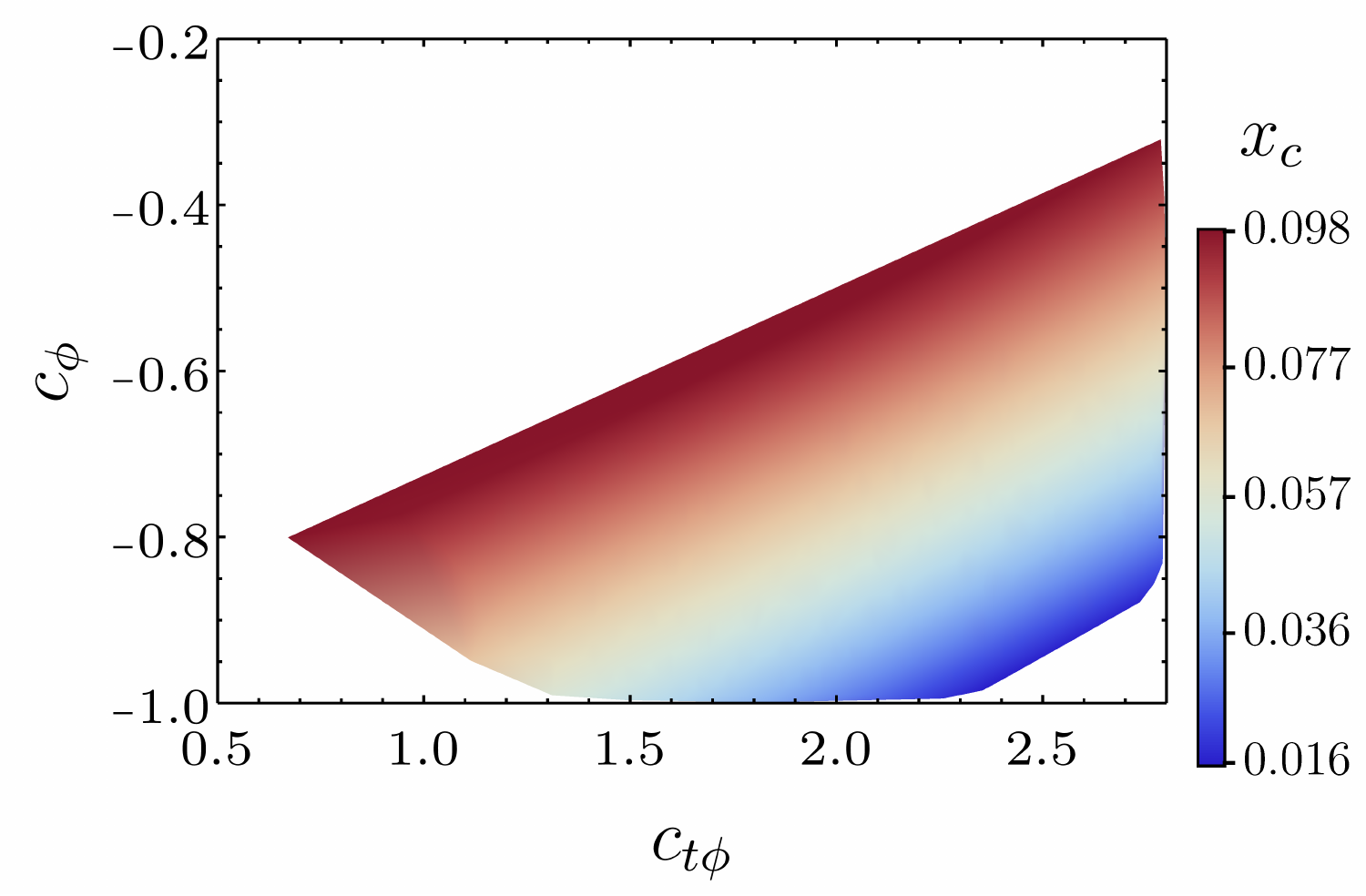}
    \caption{\it Region of the $c_{t\phi}$--$c_\phi$ parameter space in units of TeV$^{-2}$ (for fixed $c_{\phi\Box} = 0$) allowed by data that gives rise to a FOPT, and the corresponding value of $x_c$ where the SMEFT evolution curve crosses the FOPT boundary, as extracted from Ref.~\cite{Gould:2022ran}. The effect of varying $c_{\phi G}$ is negligible on the phase diagram, and it only influences the allowed values of $c_{t \phi}$, so we do not show it.
    }
    \label{fig:PT_1}
\end{figure}

\begin{figure}[t]
\includegraphics[width=\columnwidth]{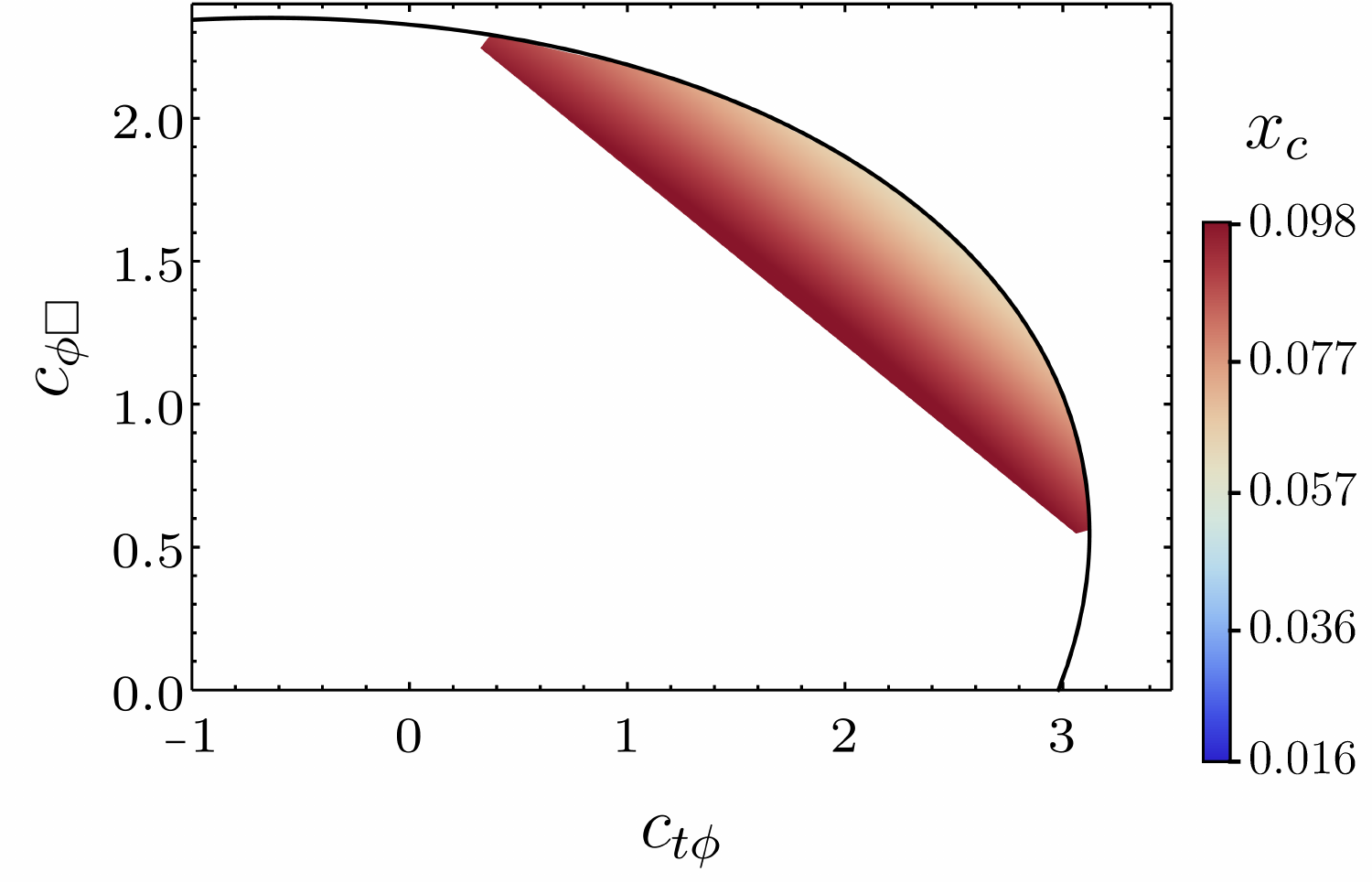}
\caption{\it Same as Fig.~\ref{fig:PT_1}, but in the $c_{t\phi}$--$c_{\phi\Box}$ plane in TeV$^{-2}$ (for fixed $c_{\phi} = 0$). 
}
\label{fig:PT_2}
\end{figure}

Relying on the phase boundary for the 3D SU(2) + Higgs theory determined using lattice methods in Ref. \cite{Gould:2022ran}, in Figs.~\ref{fig:PT_1} and \ref{fig:PT_2} we show the regions of the $c_{t \phi}$--$c_\phi$ ($c_{\phi\Box}=0$) and $c_{t \phi}$--$c_{\phi \Box}$ ($c_{\phi}=0$) planes of the parameter space, respectively, that allow for a FOPT and the corresponding value of $x=x_c$ for which it takes place. All points satisfy sanity conditions, namely that at $T=T_c$
we have $\Lambda^2 > (\pi T)^2 > \mu^2$ and $m_\phi^2/m_{W_0}^2,m_\phi^2/m_{B_0}^2 < 0.5$. Let us also note that, as discussed in Ref.~\cite{Gould:2019qek}, one can take $x > 0.01$ as an approximate regime where the inclusion of $(\phi^\dagger \phi)^3$ (as generated in the high-temperature limit of the SM) does not significantly alter the existing lattice results. 

We find remarkable that moderately positive values of $c_{t\phi}$, together with small negative values of $c_\phi$ or positive values of $c_{\phi\Box}$, drive the evolution towards the FOPT regime (by making $\lambda_3$, and thus $x$, smaller), both being allowed by experimental data~\footnote{For all points considered, $\lambda$ remains positive and (if present) $c_\phi$ is negative, so the zero-temperature potential is bounded from below at renormalizable and dimension-six level. Also, we have $\lambda_3 > 0$, so the thermal potential barrier is radiative, and not tree-level generated~\cite{Camargo-Molina:2024sde}.}. In fact, if we combine the contribution from $c_\phi$ and $c_{\phi\Box}$, together with $c_{t\phi} \sim 2.5$ TeV$^{-2}$, one can find a FOPT with both coefficients even below $\mathcal{O}(1)$ TeV$^{-2}$ in absolute value. The reason for all this is that, contrary to previous studies of the FOPT landscape of the SMEFT~\cite{Zhang:1992fs,Bodeker:2004ws,Grojean:2004xa,Ham:2004zs,Delaunay:2007wb,Grinstein:2008qi,Damgaard:2015con,Huang:2017jws,deVries:2017ncy,Cai:2017tmh,Chala:2018ari,Ellis:2018mja,Ellis:2019oqb,Ellis:2019tjf,Phong:2020ybr,Postma:2020toi,Hashino:2022ghd,Kanemura:2022txx,Alonso:2023jsi,Oikonomou:2024jms,Camargo-Molina:2021zgz,Camargo-Molina:2024sde, Hashino:2025nku}, we have accounted for all dimension-six SMEFT operators, and in particular for $\mathcal{O}_{t\phi}$, which does not contribute directly to the Higgs potential at zero temperature, but enters the matching for $\lambda_3$ relatively unsuppressed. (In Ref.~\cite{Camargo-Molina:2024sde}, the authors observed a similar trend conducting a purely perturbative study of the PT with Higgs-only dimension-six operators).
We observe how even strong PTs, necessary for successful baryogenesis and characterized by $x\lesssim 0.025$~\cite{Annala:2025aci}, are allowed for $|c_\phi| \lesssim 0.8$ TeV$^{-2}$; though they seem impossible if this WC is much closer to zero. 

\section{Conclusions}
\label{sec:discussion}
We have fully determined the dimensional reduction of the dimension-six SMEFT to $\mathcal{O}(g^4)$ accuracy, demonstrating how the relevant two-loop sum-integrals can factorize into products of one-loop sum-integrals. Using existing lattice data, our analysis exhaustively characterizes the electroweak thermal history in the regime where new physics is decoupled. In particular, we have  shown that the SMEFT can exhibit a FOPT even without large modifications to the Higgs potential at zero temperature. Both one-loop and two-loop (for the square mass terms) matching corrections have been indispensable to arrive at this conclusion. 

This finding opens new avenues in model-building~\footnote{For instance, a possible UV completion that generates $c_{t\phi}$ consists of heavy vector-like quarks with quantum numbers $(\textbf{3}, \textbf{2})_{1/3}$ that couple only to the third family quarks through a mixed Yukawa term. (Scalar operators would come from loops enhanced by sizable Yukawas among the Higgs and the new fermions, or by heavy scalars.). See e.g. Ref. \cite{Huang:2017jws}.}, motivating the search for collider signatures beyond the conventional targets of new light scalars or enhanced Higgs pair production~\cite{No:2013wsa,Huang:2017jws,Ramsey-Musolf:2019lsf,Li:2019tfd,Basler:2019nas,Biermann:2024oyy}. 
On a different front, our results provide the strongest evidence that, if $c_\phi$ is constrained to be very small, the observation of a strong EWPT (e.g. through the detection of the corresponding gravitational-wave background) would necessarily imply the failing of the SMEFT, and so hint to light new physics.

\section*{Acknowledgments} 
We thank Djuna Croon and José Santiago for useful comments. MC is supported by the European Research Council under grant agreement n. 101230200. MCF is supported by the Spanish Research Agency under grant EUR2024-153549. LG is supported by the FPU program under grant number FPU23/02026. This work has received further funding from MICIU/AEI/10.13039/501100011033 (grants PID2022-139466NB-C21/C22 and PID2024-161668NB-100) as well as from Junta de Andaluc\'ia (grants FQM 101 and P21-00199)

\appendix

\section{Tables of operators}
\label{app:operators}

In this appendix we collect all non-baryon-non-lepton-violating dimension-six operators of the SMEFT in the physical basis. They are shown in Tabs.~\ref{tab:bosonicoperators}, \ref{tab:2fermionoperators} and \ref{tab:4fermionoperators}. Pauli matrices are written as $\sigma^I$ and $\sigma_{\mu\nu} \equiv \frac{i}{2} [\gamma_\mu, \gamma_\nu]$. We note that all operators denoted as $\mathcal{O}_{f X}$ and $\mathcal{O}_{f \phi}$, where $f=u,d,e$ and $X=B,W,G$, are complex-valued.

\begin{table}[t]
 \begin{tabular}{|p{0.3\columnwidth} m{0.4\columnwidth}|}
 \hline
 \rowcolor{gray!20} \multicolumn{2}{|c|}{$\boldsymbol{X^3}$} \\
 \hline
 \hspace{0.1\columnwidth} \textcolor{gray}{$\mathcal{O}_{3G}$} & \textcolor{gray}{$f^{ABC} G_\mu^{A\nu}G_\nu^{B\rho} G_\rho^{C\mu}$} \\
 \hspace{0.1\columnwidth} \textcolor{gray}{$\mathcal{O}_{3\widetilde{G}}$} &  \textcolor{gray}{$f^{ABC} \widetilde{G}_\mu^{A\nu}G_\nu^{B\rho} G_\rho^{C\mu}$} \\
 \hspace{0.1\columnwidth} \textcolor{gray}{$\mathcal{O}_{3W}$} & \textcolor{gray}{$\epsilon^{IJK} W_\mu^{I\nu}W_\nu^{J\rho} W_\rho^{K\mu}$} \\
 \hspace{0.1\columnwidth} \textcolor{gray}{$\mathcal{O}_{3\widetilde{W}}$} &  \textcolor{gray}{$\epsilon^{IJK} \widetilde{W}_\mu^{I\nu}W_\nu^{J\rho} W_\rho^{K\mu}$} \\
 \hline
 \rowcolor{gray!20} \multicolumn{2}{|c|}{$\boldsymbol{X^2 \phi^4}$} \\
 \hline
 \hspace{0.1\columnwidth} \textcolor{gray}{$\mathcal{O}_{\phi G}$} & \textcolor{gray}{$(\phi^\dagger\phi) \, G_{\mu\nu}^A G^{A\mu\nu}$} \\
 \hspace{0.1\columnwidth} \textcolor{gray}{$\mathcal{O}_{\phi \widetilde{G}}$} & \textcolor{gray}{$(\phi^\dagger\phi) \, \widetilde{G}_{\mu\nu}^A G^{A\mu\nu}$} \\
 \hspace{0.1\columnwidth} \textcolor{gray}{$\mathcal{O}_{\phi W}$} & \textcolor{gray}{$(\phi^\dagger\phi) \, W_{\mu\nu}^I W^{I\mu\nu}$} \\
 \hspace{0.1\columnwidth} \textcolor{gray}{$\mathcal{O}_{\phi \widetilde{W}}$} & \textcolor{gray}{$(\phi^\dagger\phi) \, \widetilde{W}_{\mu\nu}^I W^{I\mu\nu}$} \\
 \hspace{0.1\columnwidth} \textcolor{gray}{$\mathcal{O}_{\phi B}$} & \textcolor{gray}{$(\phi^\dagger\phi) \, B_{\mu\nu} B^{\mu\nu}$} \\
 \hspace{0.1\columnwidth} \textcolor{gray}{$\mathcal{O}_{\phi \widetilde{B}}$} & \textcolor{gray}{$(\phi^\dagger\phi) \, \widetilde{B}_{\mu\nu} B^{\mu\nu}$} \\ 
 \hspace{0.1\columnwidth} \textcolor{gray}{$\mathcal{O}_{\phi WB}$} & \textcolor{gray}{$(\phi^\dagger\sigma^I\phi) \, W^I_{\mu\nu} B^{\mu\nu}$} \\
 \hspace{0.1\columnwidth} \textcolor{gray}{$\mathcal{O}_{\phi \widetilde{W}B}$} & \textcolor{gray}{$(\phi^\dagger\sigma^I\phi) \, \widetilde{W}^I_{\mu\nu} B^{\mu\nu}$} \\ 
 \hline
 \rowcolor{gray!20} \multicolumn{2}{|c|}{$\boldsymbol{\phi^4 D^2}$} \\
 \hline
 \hspace{0.1\columnwidth} $\mathcal{O}_{\phi\square}$ & $(\phi^\dagger\phi) \square(\phi^\dagger\phi)$\\
 \hspace{0.1\columnwidth} $\mathcal{O}_{\phi D}$ & $(\phi^\dagger D^\mu\phi)^\dagger \phi^\dagger D_\mu\phi$\\
 \hline
 \rowcolor{gray!20} \multicolumn{2}{|c|}{$\boldsymbol{\phi^6}$} \\
 \hline
 \hspace{0.1\columnwidth} $\mathcal{O}_\phi$ & $(\phi^\dagger\phi)^3$\\
 \hline
 \end{tabular}
 \caption{\it Bosonic dimension-six SMEFT operators. Terms in gray arise at loop level in weakly-coupled UV completions.}\label{tab:bosonicoperators}
\end{table}

\begin{table}[t]
\begin{tabular}{|p{0.3\columnwidth} p{0.4\columnwidth}|}
 \hline
 \rowcolor{gray!20} \multicolumn{2}{|c|}{$\boldsymbol{\psi^2 X \phi}$} \\
 \hline
 \hspace{0.1\columnwidth} \textcolor{gray}{$\mathcal{O}_{uG}$} & \textcolor{gray}{$(\overline{q} \, T^A \sigma^{\mu\nu} u) \widetilde{\phi} G_{\mu\nu}^A$}\\
 \hspace{0.1\columnwidth} \textcolor{gray}{$\mathcal{O}_{uW}$} & \textcolor{gray}{$(\overline{q}\sigma^{\mu\nu} u) \sigma^I \widetilde{\phi} W_{\mu\nu}^I$}\\
 \hspace{0.1\columnwidth} \textcolor{gray}{$\mathcal{O}_{uB}$} & \textcolor{gray}{$(\overline{q}\sigma^{\mu\nu} u) \widetilde{\phi} B_{\mu\nu}^I$}\\
 \hspace{0.1\columnwidth} \textcolor{gray}{$\mathcal{O}_{dG}$} & \textcolor{gray}{$(\overline{q} \, T^A \sigma^{\mu\nu} d) \phi G_{\mu\nu}^A$}\\
 \hspace{0.1\columnwidth} \textcolor{gray}{$\mathcal{O}_{dW}$} & \textcolor{gray}{$(\overline{q}\sigma^{\mu\nu} d) \sigma^I \phi W_{\mu\nu}^I$}\\
 \hspace{0.1\columnwidth} \textcolor{gray}{$\mathcal{O}_{dB}$} & \textcolor{gray}{$(\overline{q}\sigma^{\mu\nu} d) \phi B_{\mu\nu}^I$}\\
 \hspace{0.1\columnwidth} \textcolor{gray}{$\mathcal{O}_{eW}$} & \textcolor{gray}{$(\overline{l}\sigma^{\mu\nu} e) \sigma^I \phi W_{\mu\nu}^I$}\\
 \hspace{0.1\columnwidth} \textcolor{gray}{$\mathcal{O}_{eB}$} & \textcolor{gray}{$(\overline{l}\sigma^{\mu\nu} e) \phi B_{\mu\nu}$}\\
 \hline
 \rowcolor{gray!20} \multicolumn{2}{|c|}{$\boldsymbol{\psi^2 \phi^2}$}\\
 \hline
 \hspace{0.1\columnwidth} $\mathcal{O}_{\phi q}^{(1)}$ & $\overline{q}\gamma^\mu q (\phi^\dagger\ii \overleftrightarrow{D}_\mu\phi)$\\
 \hspace{0.1\columnwidth} $\mathcal{O}_{\phi q}^{(3)}$ & $\overline{q}\sigma^I\gamma^\mu q (\phi^\dagger\ii \overleftrightarrow{D}_\mu^I\phi)$\\
 \hspace{0.1\columnwidth} $\mathcal{O}_{\phi u}$ & $\overline{u}\gamma^\mu u (\phi^\dagger\ii \overleftrightarrow{D}_\mu\phi)$\\
 \hspace{0.1\columnwidth} $\mathcal{O}_{\phi d}$  & $\overline{d}\gamma^\mu d (\phi^\dagger\ii \overleftrightarrow{D}_\mu\phi)$\\
 \hspace{0.1\columnwidth} $\mathcal{O}_{\phi ud}$ & $\overline{u}\gamma^\mu d (\widetilde{\phi}^\dagger\ii D_\mu\phi)$\\
 \hspace{0.1\columnwidth} $\mathcal{O}_{\phi l}^{(1)}$ & $\overline{l}\gamma^\mu l (\phi^\dagger\ii \overleftrightarrow{D}_\mu\phi)$\\
 \hspace{0.1\columnwidth} $\mathcal{O}_{\phi l}^{(3)}$ & $\overline{l}\sigma^I\gamma^\mu l (\phi^\dagger\ii \overleftrightarrow{D}_\mu^I\phi)$\\
 \hspace{0.1\columnwidth} $\mathcal{O}_{\phi e}$ & $\overline{e}\gamma^\mu e (\phi^\dagger\ii \overleftrightarrow{D}_\mu\phi)$\\
 \hline
 \rowcolor{gray!20} \multicolumn{2}{|c|}{$\boldsymbol{\psi^2 \phi^3}$}\\
 \hline
 \hspace{0.1\columnwidth} $\mathcal{O}_{u\phi}$ & $\overline{q}\widetilde{\phi}u (\phi^\dagger\phi)$ \\
 \hspace{0.1\columnwidth} $\mathcal{O}_{d\phi}$ & $\overline{q}\phi d (\phi^\dagger\phi)$ \\
 \hspace{0.1\columnwidth} $\mathcal{O}_{e\phi}$ & $\overline{q}\phi e (\phi^\dagger\phi)$ \\
 \hline
\end{tabular}
\caption{\it Two-fermion dimension-six SMEFT operators. Terms in gray arise at loop level in UV completions.} \label{tab:2fermionoperators}
\end{table}

\begin{table}[t]
\begin{tabular}{|p{0.3\columnwidth} p{0.4\columnwidth}|}
 \hline
 \rowcolor{gray!20} \multicolumn{2}{|c|}{$\boldsymbol{q^4}$} \\
 \hline
 \hspace{0.1\columnwidth} $\mathcal{O}_{qq}^{(1)}$ & $(\overline{q}\gamma^\mu q) (\overline{q}\gamma_\mu q)$\\
 \hspace{0.1\columnwidth} $\mathcal{O}_{qq}^{(3)}$ & $(\overline{q}\gamma^\mu\sigma^I q) (\overline{q}\gamma_\mu\sigma^I q)$\\
 \hspace{0.1\columnwidth} $\mathcal{O}_{uu}$ & $(\overline{u}\gamma^\mu u) (\overline{u}\gamma_\mu u)$\\
 \hspace{0.1\columnwidth} $\mathcal{O}_{dd}$ & $(\overline{d}\gamma^\mu d) (\overline{d}\gamma_\mu d)$\\
 \hspace{0.1\columnwidth} $\mathcal{O}_{ud}^{(1)}$ & $(\overline{u}\gamma^\mu u) (\overline{d}\gamma_\mu d)$\\
 \hspace{0.1\columnwidth} $\mathcal{O}_{ud}^{(8)}$ & $(\overline{u}\gamma^\mu T^A u) (\overline{d}\gamma_\mu T^A d)$\\
 \hspace{0.1\columnwidth} $\mathcal{O}_{qu}^{(1)}$ & $(\overline{q}\gamma^\mu q) (\overline{u}\gamma_\mu u)$\\
 \hspace{0.1\columnwidth} $\mathcal{O}_{qu}^{(8)}$ & $(\overline{q}\gamma^\mu T^A q) (\overline{u}\gamma_\mu T^A u)$\\
 \hspace{0.1\columnwidth} $\mathcal{O}_{qd}^{(1)}$ & $(\overline{q}\gamma^\mu q) (\overline{d}\gamma_\mu d)$\\
 \hspace{0.1\columnwidth} $\mathcal{O}_{qd}^{(8)}$ & $(\overline{q}\gamma^\mu T^A q) (\overline{d}\gamma_\mu T^A d)$\\
 \hspace{0.1\columnwidth} $\mathcal{O}_{quqd}^{(1)}$ & $(\overline{q}u)\epsilon (\overline{q} d)$\\
 \hspace{0.1\columnwidth} $\mathcal{O}_{quqd}^{(8)}$ & $(\overline{q}T^A u)\epsilon (\overline{q} T^A d)$\\
 \hline
 \rowcolor{gray!20} \multicolumn{2}{|c|}{$\boldsymbol{\ell^4}$}\\
 \hline
 \hspace{0.1\columnwidth} $\mathcal{O}_{ll}$ & $(\overline{l}\gamma^\mu l) (\overline{l}\gamma_\mu l)$\\
 \hspace{0.1\columnwidth} $\mathcal{O}_{ee}$ & $(\overline{e}\gamma^\mu e) (\overline{e}\gamma_\mu e)$\\
 \hspace{0.1\columnwidth} $\mathcal{O}_{le}$ & $(\overline{l}\gamma^\mu l) (\overline{e}\gamma_\mu e)$\\
 \hline
 \rowcolor{gray!20} \multicolumn{2}{|c|}{$\boldsymbol{q^2 \ell^2}$}\\
 \hline
 \hspace{0.1\columnwidth} $\mathcal{O}_{lq}^{(1)}$ & $(\overline{l}\gamma^\mu l) (\overline{q}\gamma_\mu q)$\\
 \hspace{0.1\columnwidth} $\mathcal{O}_{lq}^{(3)}$ & $(\overline{l}\gamma^\mu\sigma^I l) (\overline{q}\gamma_\mu\sigma^I q)$\\
 \hspace{0.1\columnwidth} $\mathcal{O}_{eu}$ & $(\overline{e}\gamma^\mu e) (\overline{u}\gamma_\mu u)$\\
 \hspace{0.1\columnwidth} $\mathcal{O}_{ed}$ & $(\overline{e}\gamma^\mu e) (\overline{d}\gamma_\mu d)$\\
 \hspace{0.1\columnwidth} $\mathcal{O}_{qe}$ & $(\overline{q}\gamma^\mu q) (\overline{e}\gamma_\mu e)$\\
 \hspace{0.1\columnwidth} $\mathcal{O}_{lu}$ & $(\overline{l}\gamma^\mu l) (\overline{u}\gamma_\mu u)$\\
 \hspace{0.1\columnwidth} $\mathcal{O}_{ld}$ & $(\overline{l}\gamma^\mu l) (\overline{d}\gamma_\mu d)$\\
 \hspace{0.1\columnwidth} $\mathcal{O}_{ledq}$ & $(\overline{l} e) (\overline{d} q)$\\
 \hspace{0.1\columnwidth} $\mathcal{O}_{lequ}^{(1)}$ & $(\overline{l} e)\epsilon (\overline{q} u)$\\
 \hspace{0.1\columnwidth} $\mathcal{O}_{lequ}^{(3)}$ & $(\overline{l} \sigma^{\mu\nu} e) (\overline{q}\sigma_{\mu\nu} u)$\\
 \hline
\end{tabular}
\caption{\it Four-fermion dimension-six SMEFT operators.}\label{tab:4fermionoperators}
\end{table}

\section{Constraints on relevant SMEFT operators}
\label{app:constraints}

\begin{figure}[t]
\includegraphics[width=0.7\columnwidth]{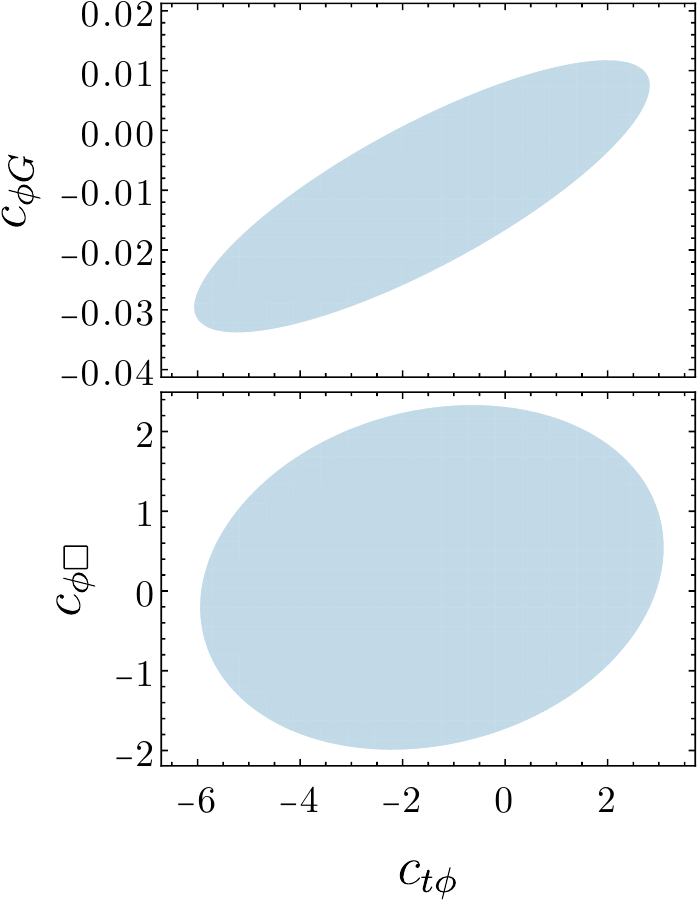}
\caption{\it Experimentally allowed values for $c_{t \phi}$ and $c_{\phi G}$, and for $c_{t \phi}$ and $c_{\phi \Box}$ in TeV$^{-2}$ from top and Higgs physics~\cite{Ellis:2020unq,Ethier:2021bye,deBlas:2025xhe}, as derived using \texttt{SMEFiT} upon marginalizing over other operators.
}
\label{fig:allowed_region}
\end{figure}

For our analysis, we mainly focus on operators which can shift $\lambda_3$ towards smaller or more negative values.  As indicated in the main text, these are $\mathcal{O}_{t \phi}$, $\mathcal{O}_{\phi G}$, $\mathcal{O}_\phi$ and $\mathcal{O}_{\phi \Box}$. In Fig.~\ref{fig:allowed_region}, we show the constraints, from \texttt{SMEFiT}, on different pairs of these SMEFT operators upon marginalization on those others that are not relevant for the analysis of the electroweak phase transition.

\section{Illustration of the phase evolution of the SM(EFT)}
\label{app:illustration}

\begin{figure}[h]
\includegraphics[width=0.8\columnwidth]{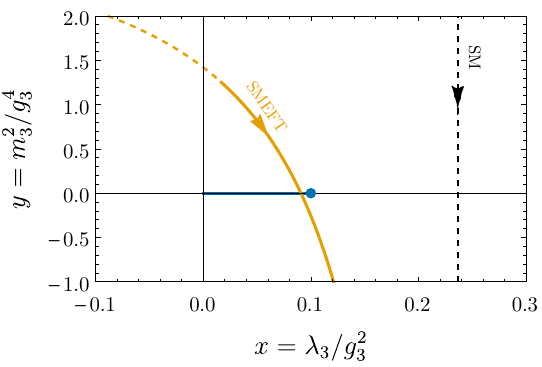}
\caption{\it Phase structure of the high-temperature limit of the SM and the SMEFT (for $\delta m_3^2=-2$, $\delta\lambda_3=-5.5$). The dot at the end of the blue line indicates a critical point. Transitions across this line are first-order.}\label{fig:phasediagram_1}
\end{figure}

We can roughly understand the evolution of the SM and the SMEFT across their phase diagram using  $\mathcal{O}(g^2)$ matching in the limit of vanishing $g'$, together with generic deformations due to SMEFT interactions:
\begin{align}
    m_3^2 &= -\mu^2+\frac{1}{4}(|Y_t|^2 + 3 g^2 + 8\lambda) T^2 + \delta m_3^2 \frac{T^4}{\Lambda^2}\,,\nonumber\\
    \lambda_3 &= \lambda T + \delta\lambda_3 \frac{T^3}{\Lambda^2}\,,\nonumber\\
    g_3 &= g \sqrt{T}\,.
\end{align}
Thus, for example, for $\delta m_3^2 = -2$, $\delta\lambda_3 = -5.5$, we obtain the phase evolution in Fig.~\ref{fig:phasediagram_1}, where we vaguely approximate the first-order phase transition boundary with a straight segment $y=0$ (corresponding to the condition that $m_3^2(T_c)=0$). We observe the well-known fact that the SM ($\delta m_3^2=\delta\lambda_3=0$) shows a crossover, but for the SMEFT there can be a FOPT.

We assume $\Lambda=1$ TeV, and the arrows in the figure indicate the temperature evolution from $T_i=\Lambda$ down to $T_f=10$ GeV. The dashed region in the SMEFT line represents the temperatures at which the SMEFT is no longer a good approximation to the UV physics, $T\gtrsim \Lambda/(2\pi)$. 

%%%%%%%%%%%%%%%%%%%%%%%%%%%%%%
\bibliographystyle{apsrev4-2}
\bibliography{refs}
%%%%%%%%%%%%%%%%%%%%%%%%%%%%%%

%%%%%%%%%%%%%%%%%%%%%%%%%%%%%%
\end{document}